\documentclass[pdflatex,aps,prl,twocolumn,nofootinbib,showpacs,superscriptaddress]{revtex4-1}
\usepackage{graphicx}
\usepackage{subfigure}
\newcommand{\nc}[1]{\newcommand{#1}}
\newcommand{\be}{\begin{eqnarray}}
\newcommand{\ee}{\end{eqnarray}}

\nc{\hmu}{\hat{\mu}}
\nc{\bmu}{\bar{\mu}}
\def\lsim{\raise0.3ex\hbox{$<$\kern-0.75em\raise-1.1ex\hbox{$\sim$}}}
\def\gsim{\raise0.3ex\hbox{$>$\kern-0.75em\raise-1.1ex\hbox{$\sim$}}}
\usepackage{color}
\usepackage{epsfig}
\usepackage{graphics}
\usepackage{amssymb}
\usepackage{soul}
\setstcolor{red}
\sethlcolor{yellow}
\nc{\fb}{\color{blue}}
\nc{\fr}{\color{red}}

\begin{document}

\title{Freeze-out Conditions in Heavy Ion Collisions from QCD Thermodynamics}
\author{
A. Bazavov}
\author{
H.-T. Ding}
\author{
P. Hegde} 
\affiliation{
Physics Department, Brookhaven National Laboratory, Upton, NY 11973, USA}
\author{
O. Kaczmarek}
\affiliation{
Fakult\"at f\"ur Physik, Universit\"at Bielefeld, D-33615 Bielefeld,
Germany}
\author{
F. Karsch}
\affiliation{
Physics Department, Brookhaven National Laboratory, Upton, NY 11973, USA}
\affiliation{
Fakult\"at f\"ur Physik, Universit\"at Bielefeld, D-33615 Bielefeld,
Germany}
\author{
E. Laermann} 
\affiliation{
Fakult\"at f\"ur Physik, Universit\"at Bielefeld, D-33615 Bielefeld,
Germany} 
\author{
Swagato Mukherjee} 
\affiliation{
Physics Department, Brookhaven National Laboratory, Upton, NY 11973, USA}
\author{
P. Petreczky}
\affiliation{
Physics Department, Brookhaven National Laboratory, Upton, NY 11973, USA}
\author{
C. Schmidt} 
\affiliation{
Fakult\"at f\"ur Physik, Universit\"at Bielefeld, D-33615 Bielefeld,
Germany}
\author{
D. Smith}
\affiliation{
Fakult\"at f\"ur Physik, Universit\"at Bielefeld, D-33615 Bielefeld,
Germany}
\author{
W. Soeldner}
\affiliation{
Institut f\"ur Theoretische Physik, Universit\"at Regensburg,
D-93040 Regensburg, Germany}
\author{
M. Wagner}
\affiliation{
Fakult\"at f\"ur Physik, Universit\"at Bielefeld, D-33615 Bielefeld,
Germany}
\date{\today}
\begin{abstract}

We present a determination of chemical freeze-out conditions in heavy ion 
collisions based on ratios of cumulants of net 
electric  charge fluctuations. These ratios can 
reliably be calculated in lattice QCD for a wide range of chemical
potential values by using a next-to-leading
order Taylor series expansion 
around the limit of vanishing baryon, electric charge and strangeness 
chemical potentials. 
From a computation of up to fourth order cumulants and charge 
correlations
we first determine the strangeness and electric charge chemical potentials 
that characterize freeze-out conditions in a heavy ion collision
and confirm that 
in the temperature range $150\ {\rm MeV} \le T \le 170\ {\rm MeV}$
the hadron resonance gas model 
provides good approximations for these 
parameters that agree with QCD calculations on the (5-15)\% level.
We then show that a comparison of
lattice QCD results for ratios of up to third order
cumulants of electric charge fluctuations
with experimental results allows to extract the freeze-out
baryon chemical potential and the freeze-out temperature.
\end{abstract}

\maketitle

\noindent
{\it 1)~Introduction:}
A central goal of experiments at the Relativistic Heavy Ion Collider 
(RHIC) \cite{RHIC} is the
exploration of the phase diagram of Quantum Chromodynamics (QCD)
at non-zero temperature ($T$)
and baryon chemical potential ($\mu_B$). In particular, a systematic
Beam Energy Scan (BES) is being performed at RHIC in order 
to search for evidence for or against the existence 
of the QCD critical point, a second order phase transition
point, that has been postulated to exist at non-vanishing baryon chemical
potential in the $T$-$\mu_B$ phase diagram of QCD \cite{CEP,CP}. 
It would be the endpoint of a line of first order phase transitions 
which then would exist for larger $\mu_B$. The measurement of
fluctuations of conserved charges, e.g. net baryon number, electric
charge and strangeness \cite{STAR,PHENIX,ALICE}, plays a 
crucial role  \cite{Ejiri} in this search for critical behavior and 
the exploration of the QCD phase diagram in general. 

Fluctuations of conserved charges generated in a heavy ion 
collision experiment may reflect thermal conditions at the time where 
the expanding medium, created in these collisions, cooled down and
diluted sufficiently so that hadrons form again. It may be questioned
whether the thermal medium at this time is in equilibrium and whether
hadronization of all species takes place  at the same time. However, 
statistical hadronization models, based on thermal
hadron distributions given by the
Hadron Resonance Gas (HRG) model, describe the hadronization
process quite successfully \cite{HRG}.
Moreover, HRG model calculations of net baryon
number fluctuations \cite{Redlich} describe well experimental 
data on net proton fluctuations \cite{STAR}.
This seems to suggest that at the time of chemical freeze-out the system
can be described by thermodynamics characterized by a  
temperature $T_f$ and a baryon chemical potential $\mu_B^f$.

The measurement of conserved net charge fluctuations can 
provide a sensitive probe for critical behavior in hot and dense 
nuclear matter only when these fluctuations are generated at a point 
in the QCD phase diagram, characterized by ($T_f,\ \mu_B^f$), that is close
to the QCD transition line and eventually is also close to the elusive
critical point. Lattice QCD calculations
provide some information on the location of the QCD transition line in 
the $T$-$\mu_B$ plane at small values of
the baryon chemical potential \cite{curvature,Fodor_curv}. The position
of the freeze-out points in this phase diagram are usually determined
by comparing experimental data on 
multiplicities of various hadron species with the HRG model
calculation \cite{HRG,cleymans}. In order to put these parameters on 
a firm basis and compare them with the QCD transition line it is 
desirable to extract the freeze-out  parameters 
by comparing experimental data with a QCD calculation. This
requires observables which are experimentally accessible and can also
reliably be calculated in QCD. The fluctuations of conserved charges
and their higher order cumulants form such a set of observables.
While net baryon number fluctuations are experimentally
accessible only through measurements of net proton number 
fluctuations \cite{STAR}, which may cause some difficulties 
\cite{Asakawa,Bzdak},
electric charge fluctuations may be easier to analyze.
We thus will focus on the latter.
As an intermediate step one should also verify to what
extent HRG model calculations and QCD calculations of freeze-out 
parameters yield consistent results when applied to the
same set of thermal observables.

We present here a calculation of ratios of cumulants of net electric 
charge and net baryon number fluctuations that can 
be formed from the first three cumulants. They are related to
mean ($M_X$) , variance ($\sigma_X^2$) and skewness ($S_X$) of the 
corresponding charge distributions, $X=B,\ Q,\ S$ for baryon number, 
electric charge and strangeness, respectively.
These ratios can 
be compared to HRG model calculations and will be used to extract the 
freeze-out temperature and baryon chemical potential 
from corresponding experimental measurements. 

\begin{figure*}[t]
\begin{center}
\includegraphics[width=58mm,height=58mm]{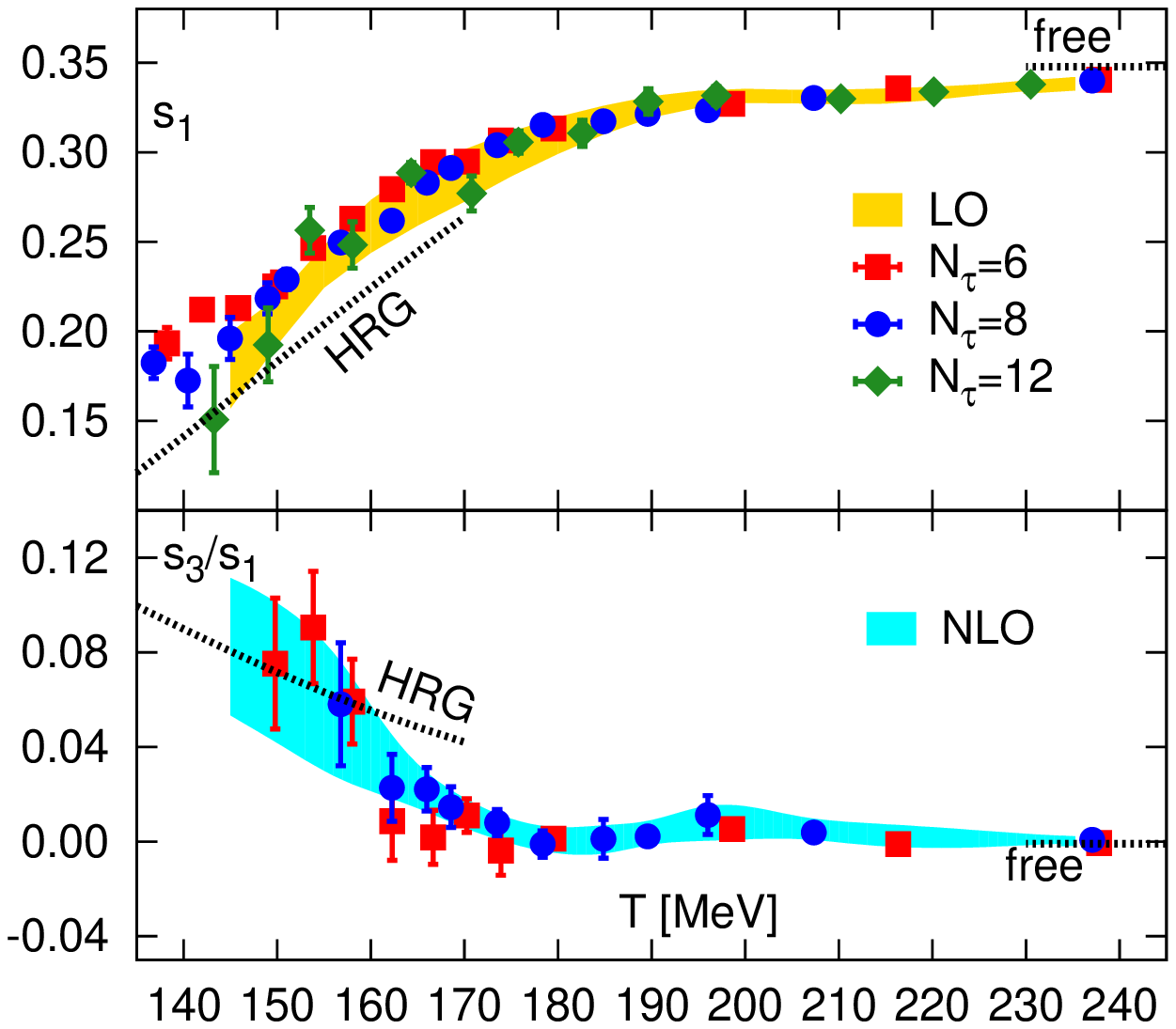} 
\includegraphics[width=58mm,height=58mm]{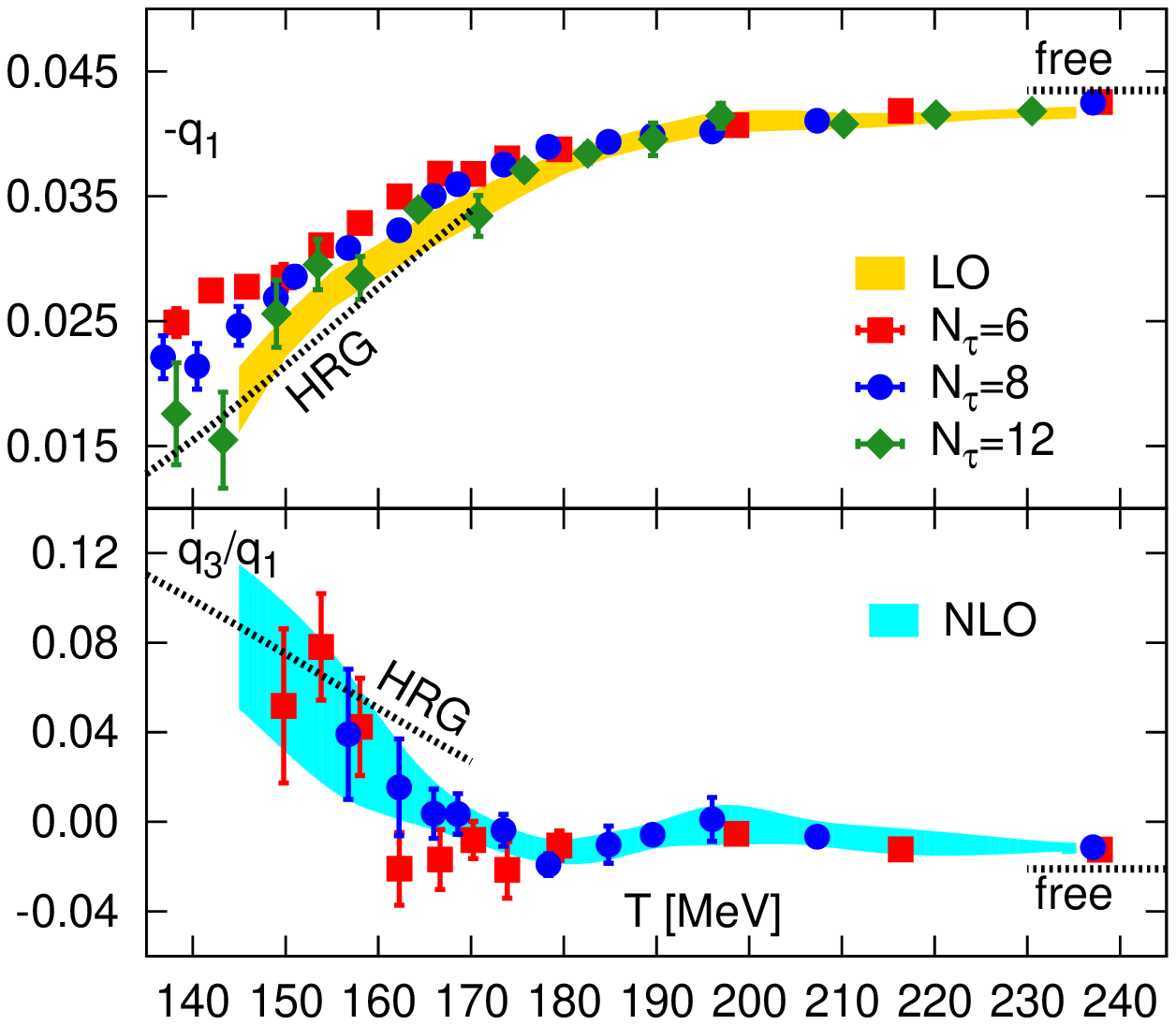} 
\includegraphics[width=58mm,height=58mm]{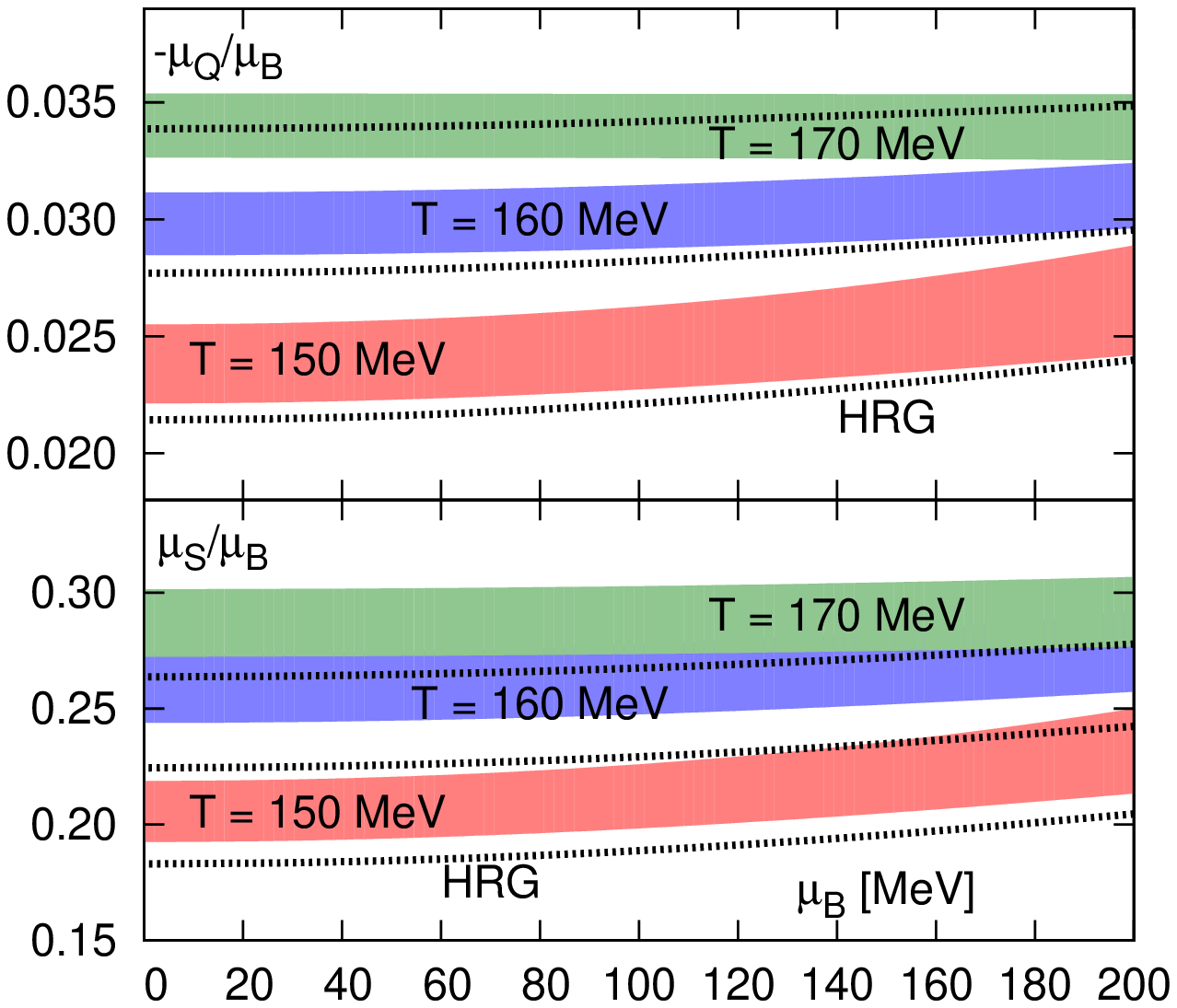} 
\caption{The leading and next-to-leading order expansion coefficients 
of the strangeness (left) and the negative of the electric charge chemical 
potentials (middle) versus temperature for $r=0.4$.
For $s_1$ and $q_1$ the LO-bands show results for the continuum 
extrapolation. For $s_3$ and $q_3$ we give an estimate for continuum 
results (NLO bands) based on spline interpolations of the $N_\tau=8$ data.
Dashed lines at low temperature are from the HRG model
and at high temperature from a massless, 3-flavor quark gas. 
The right hand panel shows NLO results for $\mu_S/\mu_B$ and
$\mu_Q/\mu_B$ as function of $\mu_B$ for three values of the 
temperature. 
\vspace*{-0.5cm}
}
\label{fig:chempot}
\end{center}
\end{figure*}

\noindent
{\it 2)~Strangeness and electric charge chemical potentials:}
In order to get access to $T_f$ and $\mu_B^f$
we need to fix the electric charge ($\mu_Q$) and strangeness ($\mu_S$) 
chemical potentials
that characterize a thermal system created in a heavy ion collision.
They are 
determined by assuming that the thermal sub-volume, probed by measuring 
fluctuations in a certain rapidity and transverse momentum window, 
reflects 
the net strangeness content and electric charge to baryon number ratios of 
the incident nuclei, 
\begin{equation}
M_S \equiv 0 \;\; ,\;\; M_Q = r M_B \;\; ,
\label{constraints}
\end{equation}
where $\displaystyle{M_X =(VT^3)^{-1}\partial \ln Z (\mu, T) /\partial 
\hmu_X}$ is the expectation value of the density 
of net charge $X$, $\mu=(\mu_B,\ \mu_Q,\ \mu_S)$ summarizes
the three charge chemical potentials
and $\hat\mu_X \equiv \mu_X/T$. At any value of ($T, \mu_B)$ 
the chemical potentials ($\mu_Q, \mu_S$) that fulfill these constraints 
can be evaluated in QCD. 
We perform a Taylor expansion of the densities
$M_X$ in terms of the three chemical potentials and calculate the expansion
coefficients of this series using the lattice regularization scheme. This 
involves the numerical calculation of generalized susceptibilities
\begin{equation}
\chi_{ijk,\mu}^{BQS} = 
\frac{1}{VT^3}\frac{\partial^{i+j+k}\ln Z (\mu, T)}{\partial\hmu_B^i\partial\hmu_Q^j
\partial\hmu_S^k}
\label{obs}
\end{equation}
at $\mu=0$. 
The calculation\footnote{In
the following subscripts and the corresponding
superscripts are suppressed in cases where the former is zero; furthermore the
abbreviation $\chi_{ijk}^{BQS}=\chi_{ijk,\mu =0}^{BQS}$ is used.}
of $\chi_{ijk}^{BQS}$ becomes computationally
demanding at higher orders, but is done in lattice QCD with steadily
increasing precision since many years 
\cite{susc_early}. In particular, recent calculations performed with
the Highly Improved Staggered Quark (HISQ) action \cite{Bazavov}
and the stout action \cite{Borsanyi}
provide continuum extrapolated results for all diagonal ($\chi_2^X$) and 
off-diagonal ($\chi_{11}^{XY}$) susceptibilities that are needed to 
determine ($\hmu_Q,\ \hmu_S$) to leading order in $\hmu_B$.

Let us write the next-to-leading order (NLO) 
expansion of $\hmu_Q$ and $\hmu_S$ as
\begin{eqnarray}
\hmu_Q = q_1\ \hmu_B + q_3\ \hmu_B^3 \; &,&\; 
\hmu_S = s_1\ \hmu_B + s_3\ \hmu_B^3  \; .
\label{mu_expansion}
\end{eqnarray}
Expanding the densities $M_X$ up to third order in the chemical potentials
we can fulfill the constraints specified in Eq.~\ref{constraints} 
at NLO.
This provides
four equations to determine the four parameters 
($s_1,\ s_3,\ q_1,\ q_3$). In leading order (LO) one obtains, 
\begin{eqnarray}
q_1 &=&
\frac{
r \left( \chi_2^B\chi_2^S - \chi_{11}^{BS}\chi_{11}^{BS} \right)
-\left( \chi_{11}^{BQ}\chi_2^S -\chi_{11}^{BS} \chi_{11}^{QS} \right)
}{
\left( \chi_2^Q\chi_2^S  - \chi_{11}^{QS} \chi_{11}^{QS} \right)
- r \left(\chi_{11}^{BQ}\chi_2^S - \chi_{11}^{BS}\chi_{11}^{QS} \right)
} 
\; ,
\nonumber \\
s_1 &=& -\frac{\chi_{11}^{BS}}{\chi_2^S} -
\frac{\chi_{11}^{QS}}{\chi_2^S} \ q_1 
\; .
\label{zeroS1}
\end{eqnarray} 
The NLO expressions are lengthy but can be derived
easily \cite{future}. We evaluated the leading order expressions in the 
temperature interval $150~{\rm MeV}\le T \le 250~{\rm MeV}$ for three 
different values of the lattice cut-off ($a$) corresponding to lattices 
with temporal extent $N_\tau \equiv 1/aT =6,\ 8$ and $12$. 
All calculations have been
performed within an ${\cal O}(a^2)$ improved gauge and staggered fermion 
(HISQ) discretization scheme \cite{hisq} for (2+1)-flavor QCD. 
The strange quark mass 
has been tuned to its physical value and the light to strange quark mass 
ratio is fixed to $m_l/m_s=1/20$, which
leads to a lightest Goldstone pion mass of about $160$~MeV. 
In the calculation of leading order results for $\hmu_S$ and $\hmu_Q$
we make use of data obtained by the HotQCD collaboration 
\cite{Bazavov}. On the $24^3\times 6$, and $32^3\times 8$ lattices
we extended these calculations in the temperature
interval $150~{\rm MeV}\le T \le 175~{\rm MeV}$ 
to 30,000 molecular dynamics time units, saving configurations
after every $10^{th}$, and 
by increasing the number of random vectors used to evaluate the 
susceptibilities to $1500$ per gauge field configuration. 

We will in the following restrict our discussion to the case, $r=0.4$, which
approximates well the situation met in Au-Au as well as Pb-Pb collisions.
The leading order expansion coefficients 
for $\hmu_Q$ and $\hmu_S$ are shown in the top panels of
Fig.~\ref{fig:chempot} left and middle.
Using spline interpolations of numerical results obtained for three 
different lattice sizes, we performed extrapolations to the continuum
limit using an ansatz linear in $1/N_\tau^2$. We have checked that 
no statistically significant differences occur by including an additional
$1/N_\tau^4$ correction. The resulting extrapolations
are shown as bands in these panels. 

In order to check the importance of NLO
corrections we have calculated $s_3$ and $q_3$ on lattices
with temporal extent $N_\tau=6$ and $8$. The results, expressed in
units of the leading order terms, are also shown in Fig.~\ref{fig:chempot}.
It is obvious from this figure that NLO corrections indeed are small.
They are negligible in the
high temperature region and are below 10\% in the temperature
interval relevant for the analysis of freeze-out conditions, i.e.,
$T\simeq (160\pm 10)$~MeV.  In fact, in this temperature range the
leading order lattice QCD results deviate from HRG model
calculations expanded to the same order by less than 15\%.
The next to leading order corrections start to become smaller than the HRG 
model values for $T\gsim 160$~MeV. This further reduces the importance of
NLO corrections.
In this respect we note that in the HRG model the NLO expansion 
reproduces the full HRG result for $\hmu_Q$ and $\hmu_S$
to better than 1.0\% for all values of $\mu_B/T \le 1.3$.
Altogether, we thus expect that the NLO-truncated QCD expansion
is a good approximation
to the complete QCD results for $\hmu_Q$ and $\hmu_S$ 
for $\mu_B\le 200$~MeV.

In order to address further systematic errors we note that
in a lattice QCD study as ours which utilizes the staggered discretization 
scheme, the biggest cut-off effects at non-vanishing lattice spacing are 
generally due to so-called taste violations. These give rise to a distorted 
hadron spectrum but mainly influence the pion sector \cite{Bazavov}.
Correspondingly the electric charge susceptibilities will be the ones
most sensitive to discretization effects while the baryon and strangeness 
sectors are largely unaffected. 
At leading order these discretization effects have been eliminated by 
taking the continuum limit. At NLO taste violation effects show up in the 
electric charge sector, Fig.1(middle). However, as the corrections themselves 
are already small, we expect their influence to be small. Furthermore, the 
taste violations can be modelled within the HRG model by replacing the pion 
mass with the average, root mean-square pion mass \cite{Bazavov} of 
our lattice calculations. Results obtained with this modified spectrum 
suggest that taste violation effects are indeed negligible
in the NLO calculation of the strangeness chemical potential and lead to
at most 5\% systematic errors in $\hmu_Q$ for $\mu_B\le 200$~MeV.

Additional systematic errors arise from the 
fact that we perform calculations with degenerate
light quark masses $m_u=m_d$, as is usually done in current lattice
QCD calculations. 
As a consequence not all susceptibilities $\chi^{BQS}_{ijk}$
are independent;
actually there are two constraints in leading order
($\chi_2^B=2 \chi_{11}^{BQ}-\chi_{11}^{BS}$, 
$\chi_2^S=2 \chi_{11}^{QS}-\chi_{11}^{BS}$) and six constraints in 
next to leading order \cite{future}. These constraints of course, do not 
hold in the HRG model and may be considered as an additional source
for the distortion of the hadron spectrum. Imposing these constraints 
by hand in the HRG model
calculations we find that $q_1$ and $q_3$ can change by up to 3\% while
modifications of $s_1$ and $s_3$ are below the 1\% level. This 
suggests that even after extrapolating to the continuum limit, the 
current lattice QCD calculations of $\mu_Q/\mu_B$ do
have an inherent systematic error of about 3\%. 

Our results for the strangeness and electric charge chemical
potentials at NLO as function of $\mu_B$ and $T$
are shown in Fig.~\ref{fig:chempot}(right). While $\mu_S/\mu_B$
varies between $0.2$ and $0.3$ in the  interval
$150~{\rm MeV}\le T\le 170~{\rm MeV}$, 
the absolute value of $\mu_Q/\mu_B$ is an order of magnitude smaller.
Both ratios are almost constant for $\mu_B\le 200$~MeV, which
is consistent with HRG model calculations.

\noindent
{\it 3) Ratios of cumulants of net charge fluctuations:}
We now are prepared to evaluate cumulants of net charge fluctuations
at as function of $T$ and $\mu_B$ at
non-vanishing values for $\mu_S$ and $\mu_Q$
that obey the constraints appropriate for thermal conditions met in
a heavy ion collision, i.e., Eq.~\ref{constraints}. 
Of particular interest are ratios of cumulants, 
$R_{nm}^X=\chi_{n,\mu}^X / \chi_{m,\mu}^X$, which to a large extent
eliminate the dependence of cumulants on the freeze-out volume.
Ratios with $n+m$ even are non-zero for $\mu=0$,
while the odd-even ratios are in leading order
proportional $\hmu_B$ and thus vanish for $\mu=0$. 
Ratios with $n+m$ even or odd
thus provide complementary information on $T_f$ and $\mu_B^f$.
We will concentrate here on the simplest such ratios,

\begin{eqnarray}
\hspace*{-0.5cm}R_{12}^X &\equiv& 
\frac{M_X}{\sigma_X^2} = 
\hmu_B \left(
R_{12}^{X,1} + R_{12}^{X,3}\ \hmu_B^2 + {\cal O}(\hmu_B^4)
\right)\; ,
\label{R12} \\
\hspace*{-0.5cm}R_{31}^X &\equiv& 
\frac{S_X \sigma_X^3}{M_X} = 
 R_{31}^{X,0} + R_{31}^{X,2}\ \hmu_B^2 + {\cal O}(\hmu_B^4)\; ,
\label{R31}
\end{eqnarray}
with $X=B,\ Q$.
These ratios can be calculated
in QCD as well as in the HRG model \cite{Ejiri}, and eventually can be 
compared to experimental data in order to determine $T_f$ and $\mu_B^f$.
We evaluated them up to ${\cal O}(\hmu_B^3)$ in
a Taylor expansion for $R_{12}^X$ and to leading order for
$R_{31}^X$. 

\begin{figure}[t]
\includegraphics[width=58mm,height=55mm]{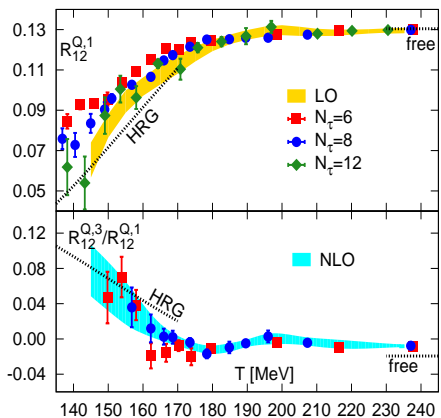}
\caption{The leading (top) and next-to-leading (bottom) order
expansion coefficients of
the ratio of first to second order cumulants of
net electric charge fluctuations versus temperature
for $r=0.4$. The bands and lines are as in 
Fig.~\protect\ref{fig:chempot}(left).
\vspace*{-0.5cm}
}
\label{fig:R12X}
\end{figure}

Let us first discuss the odd-even ratios $R_{12}^X$.
Using 
$\chi_{11}^{XX}\equiv \chi_2^X$
the LO
coefficients can be written as 
\begin{equation}
R_{12}^{X,1} =  
\frac{\chi_{11}^{BX}}{\chi_2^X}+
q_1 \frac{\chi_{11}^{XQ}}{\chi_2^X} + 
s_1 \frac{\chi_{11}^{XS}}{\chi_2^X} \; ,
\label{R12X1}
\end{equation}
with $X=B,\ Q$.
They have been evaluated
on lattices with temporal extent $N_\tau=6,\ 8$ and $12$ and have been
extrapolated to the continuum limit in the same way as for
$q_1$ and $s_1$. The LO ratio $R_{12}^{Q,1}$, evaluated for three different 
values of
the lattice cut-off (data points) and the resulting
continuum extrapolation (LO band) 
are shown in Fig.~\ref{fig:R12X}(top). In Fig.~\ref{fig:R12X}(bottom) 
we show the NLO corrections which have been evaluated on 
lattices with temporal extent $N_\tau =6$ and $8$. 
The NLO corrections to the ratio of electric charge cumulants
are below 10\%, which makes the leading order result a
good approximation for a large range of $\hmu_B$. 
Systematic errors arising from the truncation of the Taylor series for 
$R_{12}^Q$
at next-to-leading order may again be estimated by comparing the full
result in the HRG model calculation with the corresponding truncated
results. Here we find for $T=(160\pm 10)$~MeV and 
$\mu_B/T\le 1.3$ that the difference is less than 1.0\%.
Moreover, we estimated that taste violation effects in the 
NLO calculation lead to systematic errors that are at most 5\%
and thus will be negligible in $R_{12}^Q$.
Taylor series
truncated at NLO are thus expected to give a good 
approximation to the full result for a wide range of baryon
chemical potentials.
Similar results hold for the ratio $R_{12}^B$, although NLO
corrections are larger in this case.

\begin{figure*}[t]
\begin{center}
\includegraphics[width=58mm]{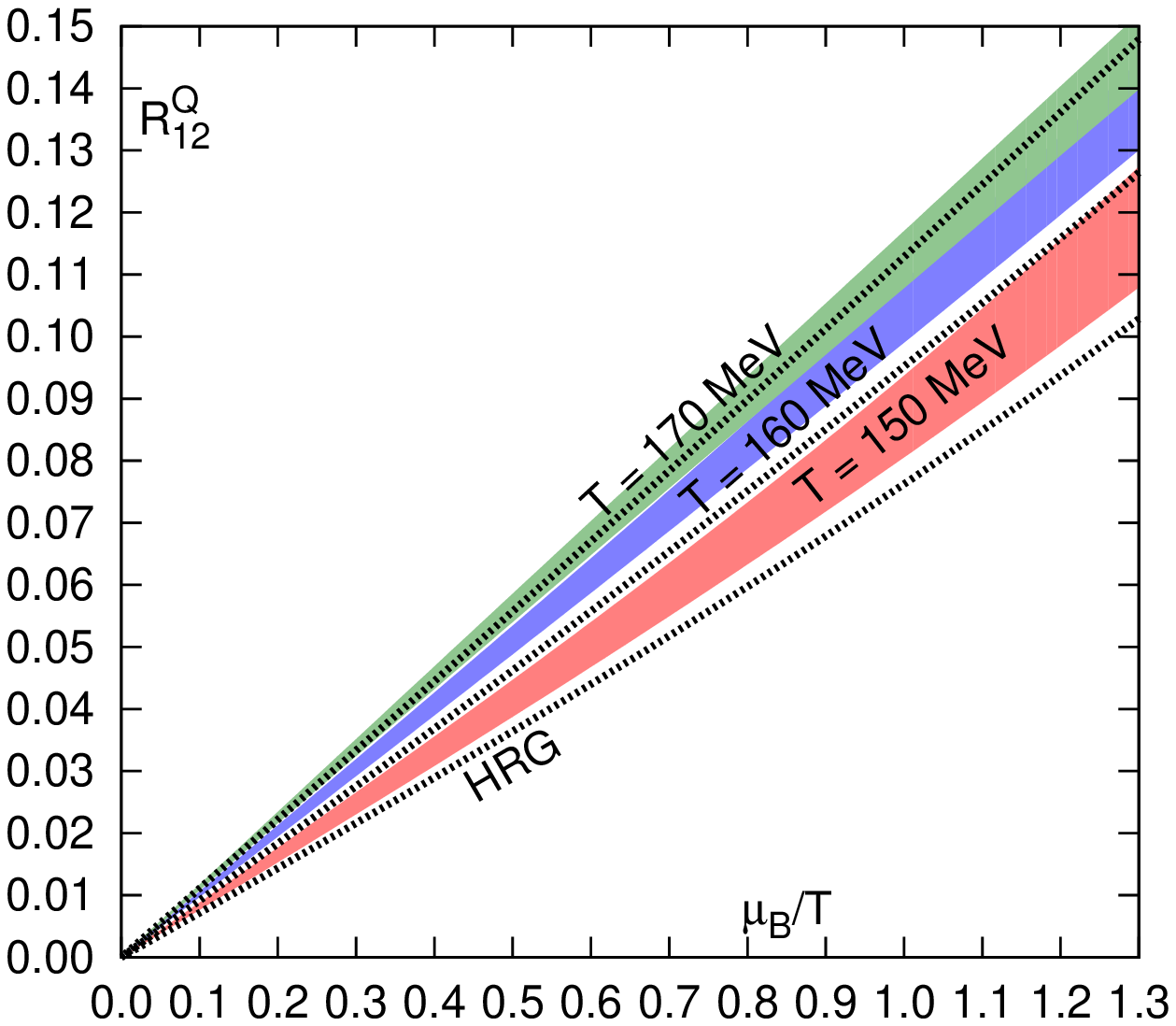}
\includegraphics[width=58mm]{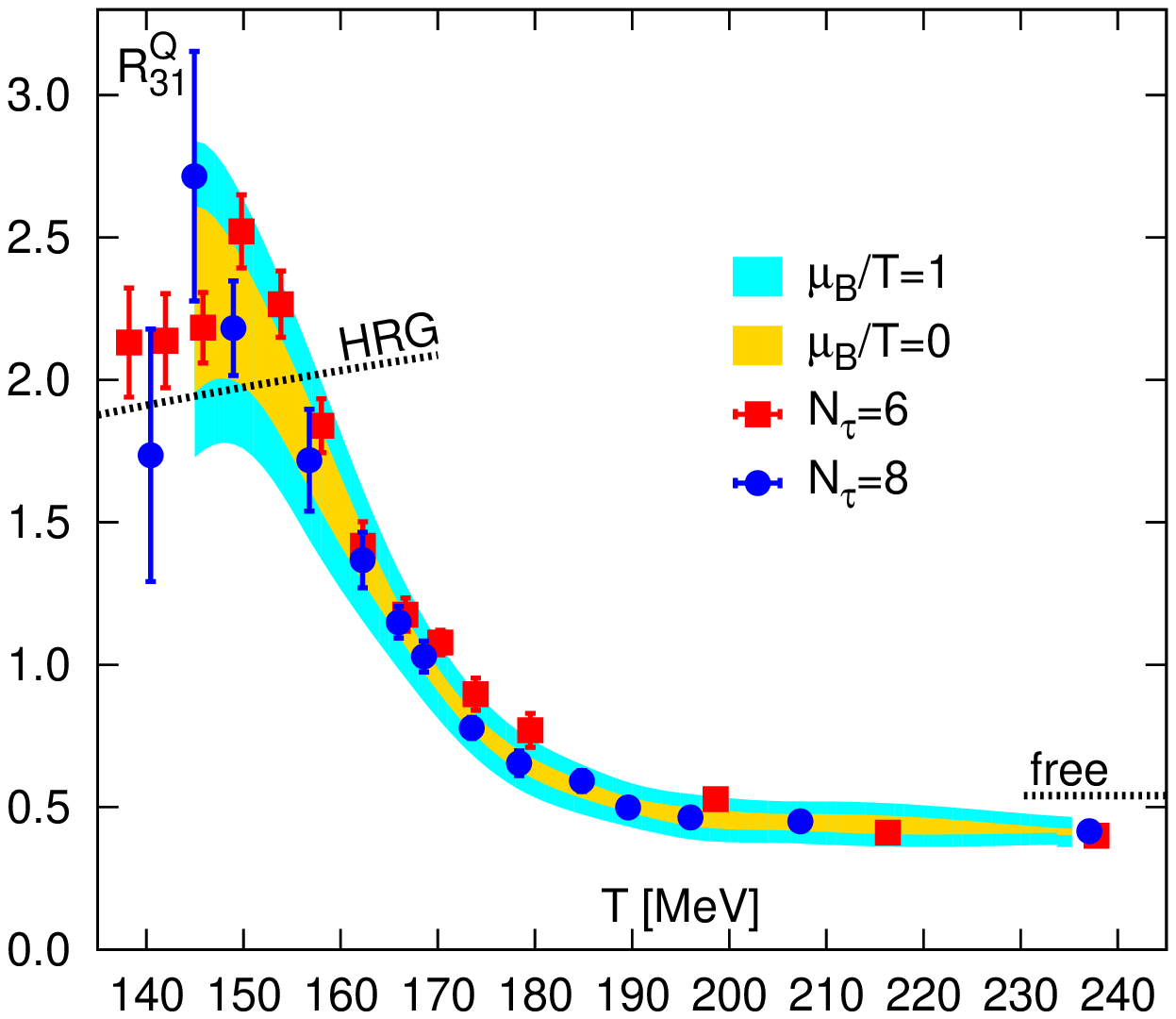}
\includegraphics[width=58mm]{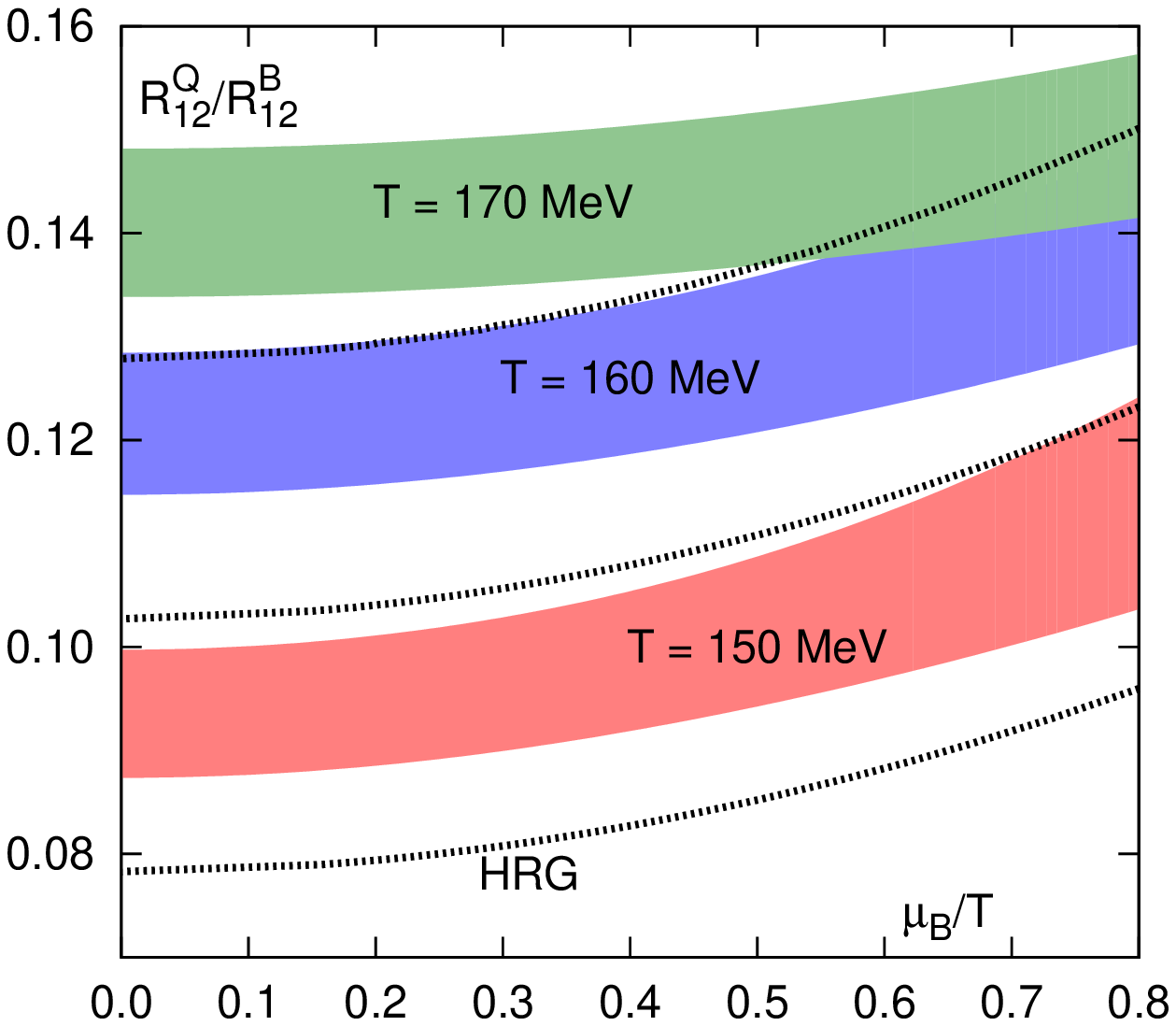}
\caption{The ratios
$R_{12}^Q$ versus $\mu_B/T$ (left) for three
values of the temperature and  $R_{31}^Q$ versus temperature
for $\mu_B=0$ (middle). The wider band on the data set for $N_\tau=8$ (middle) 
shows
an estimate of the magnitude of NLO corrections. The right hand panel shows
the NLO result for the ratio of ratios of net electric charge and baryon
number fluctuations, respectively.
}
\label{fig:R31}
\end{center}
\vspace*{-0.5cm}
\end{figure*}

\noindent
{\it 4.) Determination of freeze-out baryon chemical potential and
temperature:}
Obviously the ratio $R_{12}^Q$ shows a strong sensitivity on $\mu_B$
but varies little with $T$ for 
$T\simeq (160\pm 10)$~MeV. We show this ratio, evaluated in this
temperature interval in a NLO Taylor
expansion, in Fig.~\ref{fig:R31}(left) as function of $\hmu_B$.
For the determination of $(T_f,\mu_B^f$) a second,
complimentary information is needed. To this end we use the ratio $R_{31}^Q$, 
which is  strongly dependent on $T$ but 
receives corrections only at ${\cal O}(\hmu_B^2)$.
The leading order result for this ratio
is shown in Fig.~\ref{fig:R31}(middle).
Apparently this
ratio shows a characteristic temperature dependence for 
$T\gsim 155$~MeV that is quite different from that of HRG model calculations.
The NLO correction to this ratio vanishes in the high temperature limit
and at low $T$ the HRG model also suggests small corrections. In fact, in 
the HRG model the
LO contributions to $R_{31}^Q$ differ by less than
2\% from the exact results on the freeze-out curve for $\mu_B \le 200$~MeV.
In the transition region 
a preliminary $6^{th}$ order calculation at $T=162$~MeV \cite{future}
suggests that the $\mu_B^2$ correction in units of the LO term 
is $-0.03(10)$. Based on these estimates in the three T regions
we expect the NLO corrections to be of the order of 10\% for the whole 
temperature range.
In Fig.~\ref{fig:R31}(middle) we show the spline interpolation for the 
$N_\tau=8$ data as a band and added on top of this a band that estimates 
the effect of a 10\% contribution of the NLO correction.
The ratio $R_{31}^Q$ thus seems to be well suited for a determination of 
the freeze-out temperature. 

We now are in the position to extract $\mu_B^f$ and $T_f$ from
$R_{12}^Q$ and $R_{31}^Q$ which eventually will be measured in the 
BES at RHIC. A large value for $R_{31}^Q$, i.e. $R_{31}^Q\simeq 2$
would suggest a low freeze-out temperature $T\lsim 155$~MeV,
while a value $R_{31}^Q\simeq 1$ would suggest a large freeze-out temperature,
$T\sim 170$~MeV. A value of $R_{31}^Q\simeq 1.5$ would correspond to
$T\sim 160$~MeV. 

A measurement of $R_{31}^Q$ thus suffices to determine the freeze-out
temperature. In the HRG model parametrization of the freeze-out curve
\cite{cleymans} the favorite value for $T_f$ in the beam energy
range $200~{\rm GeV}\ge \sqrt{s_{NN}}\ge 39~{\rm GeV}$ indeed varies
by less than $2$~MeV and is about 165~MeV. At this temperature the
values for $R_{31}^Q$ calculated in the HRG model and in QCD differ quite
a bit, as is obvious from Fig.~\ref{fig:R31}. While $R_{31}^Q\simeq 2$ in
the HRG model, one finds $R_{31}^Q\simeq 1.2$ in QCD at $T=165$~MeV.
Values close to the HRG value are compatible with QCD calculations
only for $T\lsim 157$~MeV.
We thus expect to either find freeze-out temperatures that are about
5\% below HRG model results or values for $R_{31}^Q$ that are significantly
smaller than the HRG value. A measurement of this cumulant ratio at RHIC
thus will allow to determine $T_f$ and probe the consistency with HRG
model predictions.

For any of these temperature values a comparison of
an experimental value for $R_{12}^Q$ with Fig.~\ref{fig:R31}(left) will
allow to determine $\mu_B^f$. 
To be specific let us discuss the results obtained at $T=160$~MeV.
Here we find
\begin{equation}
R_{12}^Q(T=160\ {\rm MeV}) =
0.102(5)\hmu_B + 0.002(1) \hmu_B^3 \; .
\label{R12mu}
\end{equation}

For a value of the freeze-out temperature close to $T=160$~MeV
we thus 
expect to find 
$\mu_B^f=(20-30)$~MeV, if $R_{12}^Q$ lies in the range $0.012-0.020$,
$\mu_B^f=(50-70)$~MeV for $0.032\le R_{12}^Q\le 0.045$ and
$\mu_B^f=(80-120)$~MeV for $0.05\le R_{12}^Q\le 0.08$.
These parameter ranges are expected \cite{RHIC,cleymans,STARfreeze}
to cover the regions relevant for RHIC beam energies 
$\sqrt{s_{NN}}=200$~GeV, $62.4$~GeV and $39$~GeV, respectively.
As is evident from Fig.~\ref{fig:R31}(left) the values for $\mu_B^f$
will shift to smaller (larger) values when $T_f$ turns out to be 
larger (smaller) than $160$~MeV. A more refined analysis of 
$(T_f,\mu_B^f)$
will become possible, once the ratios
$R_{12}^Q$ and $R_{31}^Q$ have been measured experimentally.

\noindent
{\it 5) Conclusions:} We have shown that the first three cumulants 
of net electric charge fluctuations are well suited for
a determination of freeze-out parameters in a heavy ion collision. 
Although the ratios $R_{12}^Q$ and $R_{31}^Q$ are sufficient to
determine $T_f$ and $mu_B^f$, it will clearly be advantageous to 
have several ratios, including cumulants of net baryon number 
fluctuations, at hand that will allow to probe the 
consistency of an equilibrium thermodynamic description of cumulant
ratios at the time of freeze-out. In particular the 
ratio of ratios $R_{12}^Q/R_{12}^B = r \chi_{2,\mu}^B/\chi_{2,\mu}^Q$
is also well determined in (lattice) QCD calculations \cite{Bazavov}.
In Fig.~\ref{fig:R31}(right) we show the NLO result for this ratio
of ratios in the temperature range $T=(160\pm10)$~MeV.
Its measurement will, on the one hand, allow to probe our basic 
assumptions on constraining the electric charge and strangeness chemical 
potentials and, on the other hand, constrain possible differences
in cumulant ratios of net proton and net baryon number fluctuations. 
Once the ratios
of lower order cumulants have been used to fix the freeze-out
parameters, the calculation of higher order cumulants is parameter
free and provides unique observables for the discussion of possible
signatures for critical behavior along the freeze-out line.

{\it Acknowledgements:}
We thank 
P. Bialas, T. Luthe and L. Wresch for discussions
and help with the software development for the Bielefeld GPU cluster.
Numerical 
calculations have also been performed on the USQCD GPU-clusters
at JLab and NYBlue at the NYCCS.
This work has been supported in part by contract DE-AC02-98CH10886
with the U.S. Department of Energy, the Bundesministerium f\"ur Bildung und
Forschung under grant 06BI9001, the Gesellschaft f\"ur
Schwerionenforschung under grant BILAER, the Deutsche
Forschungsgemeinschaft under grant GRK881, and the EU Integrated
Infrastructure Initiative HadronPhysics2.


\begin{thebibliography}{99}

\bibitem{RHIC}
B.~Mohanty [STAR Collaboration],
  J.\ Phys.\ G {\bf 38}, 124023 (2011).
\bibitem{CEP}
M. Asakawa, K. Yazaki,  Nucl. Phys. A {\bf 504},  668 (1989).
\bibitem{CP}
J.~Berges, K.~Rajagopal,
  Nucl.\ Phys.\  B {\bf 538}, 215 (1999);
A.~M.~Halasz et al., 
  Phys.\ Rev.\  D {\bf 58}, 096007 (1998).
\bibitem{STAR}
  M.~M.~Aggarwal {\it et al.} [STAR Collaboration],
  Phys.\ Rev.\ Lett.\  {\bf 105}, 022302 (2010).
\bibitem{PHENIX}
A.~Adare {\it et al.}  [PHENIX Collaboration],
  Phys.\ Rev.\ C {\bf 78}, 044902 (2008).
\bibitem{ALICE}
B. Abelev {\it et al.}, [ALICE Collaboration],
  arXiv:1207.6068 [nucl-ex].
\bibitem{Ejiri}
S.~Ejiri, F.~Karsch, K.~Redlich,
  Phys.\ Lett.\ B {\bf 633}, 275 (2006).
\bibitem{HRG}
  P.~Braun-Munzinger, K.~Redlich, J.~Stachel,
  In *Hwa, R.C. (ed.) et al.: Quark gluon plasma* 491-599.
\bibitem{Redlich}
F.~Karsch and K.~Redlich,
  Phys.\ Lett.\ B {\bf 695}, 136 (2011);
P.~Braun-Munzinger {\it et al.},
  Phys.\ Rev.\ C {\bf 84}, 064911 (2011).
\bibitem{curvature}
O.~Kaczmarek {\it et al.},
  Phys.\ Rev.\  {\bf D83}, 014504 (2011).  
\bibitem{Fodor_curv}
  G.~Endrodi {\it et al.}, 
  JHEP {\bf 1104}, 001 (2011).
\bibitem{cleymans}
J.~Cleymans {\it et al.}, 
  Phys.\ Rev.\  C {\bf 73}, 034905 (2006).
\bibitem{Asakawa}
M.~Kitazawa, M.~Asakawa,
  Phys.\ Rev.\ C {\bf 85}, 021901 (2012).
\bibitem{Bzdak}
A.~Bzdak, V.~Koch, V.~Skokov,
  arXiv:1203.4529 [hep-ph].
\bibitem{susc_early}
R.~V.~Gavai, S.~Gupta,
  Phys.\ Rev.\ D {\bf 68}, 034506 (2003);
C.~R.~Allton et al.,
  Phys.\ Rev.\ D {\bf 71}, 054508 (2005);
M.~Cheng et al.,
  Phys.\ Rev.\ D {\bf 79}, 074505 (2009).
\bibitem{Bazavov} 
  A.~Bazavov {\it et al.}  [HotQCD Collaboration],
  arXiv:1203.0784 [hep-lat].
and Phys.\ Rev.\ D {\bf 85}, 054503 (2012).
\bibitem{Borsanyi} 
  S.~Borsanyi {\it et al.}, 
  JHEP {\bf 1201}, 138 (2012).
\bibitem{STARfreeze}
B.~I.~Abelev {\it et al.}  [STAR Collaboration],
  Phys.\ Rev.\ C {\bf 79}, 034909 (2009).
\bibitem{hisq}
E.~Follana {\it et al.} [HPQCD and UKQCD Collaborations],  
Phys.\ Rev.\ D {\bf 75}, 054502 (2007).
\bibitem{future}
Bielefeld-BNL Collaboration, in preparation.
\end{thebibliography}
\end{document}